\documentclass[useAMS, usenatbib]{mn2e}
\usepackage{graphicx}
\input{epsf}
\usepackage{epsfig}
\title[Eclipsing Binary Stars in M31]{A Survey of Eclipsing Binary 
Stars in the Eastern Spiral Arm of M31} 
\author[Ian Todd, Don Pollacco, Ian Skillen, 
D.M. Bramich, Steve Bell and Thomas Augusteijn]{Ian Todd$^1$\thanks{E-mail:
I.Todd@qub.ac.uk}, Don Pollacco$^1$, Ian Skillen$^2$, 
D.M. Bramich$^3$, Steve Bell$^4$ and  
\newauthor
Thomas Augusteijn$^5$\\
$^1$APS Division, Department of Physics and Astronomy, Queen's
University of Belfast, Belfast, BT7 1NN, UK\\
$^2$Isaac Newton Group of Telescopes, Apartado 321, 
E-38700 Santa Cruz de La Palma, Tenerife, Spain\\
$^3$Department of Physics and Astronomy, University of St~Andrews, 
North Haugh, Fife, KY16 9SS, UK\\
$^4$HM Nautical Almanac Office, CCLRC Rutherford Appleton
Laboratory, Chilton, Didcot, Oxon., OX11 0QX, UK \\ 
$^5$Nordic Optical Telescope, Apartado 474, 
E-38700 Santa Cruz de La Palma, Tenerife, Spain\\}

\begin{document}

\date{Accepted ????. Received ????; in original form ?????}

\pagerange{\pageref{firstpage}--\pageref{lastpage}} \pubyear{2005}

\maketitle

\label{firstpage}

\begin{abstract}
Results of an archival survey are presented using $B$-band imaging of the
eastern spiral arm of M31. Focusing on the eclipsing binary star
population, a matched-filter technique has been used to identify 280
binary systems. Of these, 127 systems (98 of which are newly
discovered) have sufficient phase coverage to allow accurate orbital
periods to be determined. At least nine of these binaries are detached 
systems which could, in principle, be used for distance determination. The
light curves of the detached and other selected systems are presented
along with a discussion of some of the more interesting binaries.
The impact of unresolved stellar blends on these light curves is
considered.
\end{abstract}

\begin{keywords}
binaries: eclipsing, galaxies: Local Group, galaxies: M31
\end{keywords}

\section{Introduction}
The determination of accurate distances to the galaxies of the Local
Group is of fundamental importance in several areas of modern astronomy. 
These galaxies serve as a natural
laboratory for studies of stellar formation and evolution over a wide
range of physical environments. A knowledge of their distances permits 
the determination of the intrinsic physical properties of their resolved
stellar content and provides an insight into population
synthesis modelling of the formation and evolution of galaxies.
Furthermore, the Local Group galaxies provide a critical step in
determining the cosmological distance scale by calibrating standard
candles such as Cepheids, extending the distance
ladder to galaxies far beyond the Local Group. This leads to a
determination of the Hubble Constant and the age of the
universe. Recent results, primarily from the Distance Scale Key
Project \citep{fm01,mould04}, have led to considerable progress in
this field and $H_{0}$ is now believed to be known to a precision of
10 per cent. At this level, the uncertainty is dominated by two systematic
effects -- the absolute distance to the LMC (the first step in the
extragalactic distance scale) and the possible dependence of the Cepheid
Period-Luminosity relationship on metallicity.

Detached eclipsing binary stars (EBs) offer the possibility of
determining the {\it absolute} properties of a stellar system such as
masses and radii.  \citet{kaluzny98} have suggested that observations
of detached EBs can be used to determine accurate absolute
luminosities and hence their distance to better than 5 per cent and
possibly even 1 per cent; see also \citet{andersen91,clausen04}. As
these distances are based mainly on geometrical arguments with only
limited physical input they are often considered to be amongst the
most reliable. Hence, distances derived from EBs can be used to
calibrate the Cepheid period-luminosity relationship. More recently it
has been proposed by \citet{wilson04} that semi-detached systems could
be used (possibly in preference to detached systems) to the same end.
While light curves from these systems may appear more complicated, the
physical processes underlying the variations ({\it e.g.}  irradiation 
and tidal effects) are now well understood.

\citet{gap68} first suggested that EBs could be used to determine
distances to the Magellanic Clouds (MCs). In fact in recent years it
has become clear that distances derived from EBs can give distances at
least comparable to that obtained from the Cepheid period-luminosity
relationship. For example, \citet{harries03} and \citet{hilditch2005}
have studied 50 SMC binaries and derived a distance of 60.6\,kpc with
an uncertainty of $\sim$5 per cent. Including results from other, equally
reliable, surveys of SMC binaries lead to an overall dispersion 
in results of $\sim$10
per cent which is most likely due to depth effects in this galaxy.

Beyond the MCs, the Andromeda Galaxy, M31, is an important distance
scale calibrator. It does not suffer from the extreme metallicities of
the MCs and, being a spiral galaxy, its geometry is far better
understood than that of the irregular MCs.  This galaxy is also a
fundamental calibrator of the zero-points of the planetary nebula and
globular cluster luminosity functions, and is the first step of the
Tully-Fisher relationship for spiral galaxies.  Consequently, M31 is a
more appropriate Local Group standard calibrator than the LMC
\citep{clem}. However, recent distance estimates for M31 based on
Cepheids and the brightness of the tip of the red giant branch ({\it e.g.},
\citep{mcc2005,fm01}) show discrepancies at the $\sim 10$ per cent level.

There have been several surveys of M31 specifically to detect variable
stars. For example, \citet{bs1963} used photographic plates to identify
684 variables in four fields, $\sim$58 per cent of which were Cepheids and
a further 9 per cent EBs.  Currently, a total of around 300
binaries are known from various surveys \citep{guinan04}.  

The DIRECT Project \citep{kaluzny98} has been attempting to derive 
distances to M31 and M33 from EBs with sufficient accuracy to calibrate 
the Cepheid distances. Using 1.0-m telescopes they have discovered a 
total of $\sim$130 EBs in both galaxies.

In this paper, we present the results of an analysis of archival
images obtained over a three year baseline and centred on the eastern
arm of M31. We use these data to identify eclipsing systems, with an 
emphasis on those suited to detailed follow-up and distance
determination.

\section{Observations and Data reduction}
The data forming the basis of this study were obtained over the period
2000-2003 by scheduled observers on the 2.5-m Isaac Newton Telescope
(INT) on the Island of La Palma in the Canary Islands. These data were 
retrieved from the Isaac Newton Group of Telescopes (ING) Archive
located at CASU, University of Cambridge.\footnote{ING archive at
http://archive.ast.cam.ac.uk}

The INT is equipped with a Wide Field Camera (WFC) comprising four
4096$\times$2048 EEV42 detectors.  The f/3.3 prime focus of the INT
has a plate scale of 0.33\,arcsec pixel$^{-1}$ on the CCD and the
camera has a relatively unvignetted field of $\sim$ 1\,000\,arcmin$^{2}$ 
across the four chips. Data were obtained in weekly runs,
scheduled in September or October of each year.  $BV$
exposures were made at approximately alternate intervals; around 200
images in $B$ and 169 in $V$ were taken, with exposure times of 900\,s
in each filter. The data used here were obtained in median 
seeing better than
1.3\,arcsec, and a summary of the images used is given in
Table~\ref{observations}.  The observed fields approximately coincide 
with those of
the DIRECT Project (fields A -- D) and the fields of \citet{mag96} in
the rich eastern spiral arm. 
The data are well sampled on short
timescales, but are less well sampled for the intermediate periods of
4-10\,d which are typical of massive, detached binaries.

\begin{figure*}
\caption{The four WFC fields surveyed in this study superimposed on a DSS2
image.}  
\includegraphics[height=10cm]{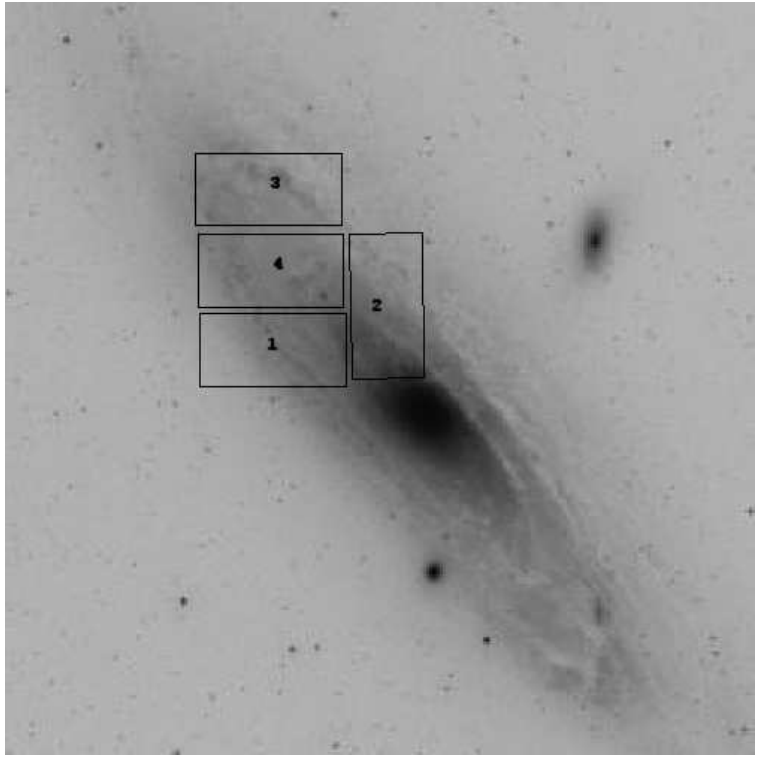}
\label{fields}
\end{figure*}

\setcounter{table}{0}
\begin{table}
\begin{minipage}{80mm}
\centering
\caption{Summary of observation log. The precise number of observations 
for each star will vary depending on the quality of its point spread
function on that particular image, {\it e.g.} being near the edge of, or near
a bad column on the CCD.}
\label{observations}
\begin{tabular}{@{}ccccc}
\hline
Year & \# B & \# V & HJD Range& Median Image \\
     & Images & Images  & 2,450,000+ & Quality (arcsec)\\
\hline
2000 & 32 & 30 & 1811.5 -- 1814.5 & 0.99 \\
2001 & 35 & 37 & 2200.4 -- 2206.5 & 1.20 \\
2002 & 52 & 18 & 2546.4 -- 2550.5 & 1.30 \\
2003 & 81 & 84 & 2906.4 -- 2910.7 & 0.97 \\
\hline
\end{tabular}
\end{minipage}
\end{table}

\subsection{Data reduction}
Each of the four fields comprising the WFC was analysed individually.
Reduction to science frames proceeded automatically using the
IRAF\footnote{IRAF is written and supported by the IRAF programming
group at the National Optical Astronomy Observatories (NOAO) in
Tucson, Arizona. NOAO is operated by the Association of Universities
for Research in Astronomy (AURA), Inc. under co-operative agreement
with the National Science Foundation} data reduction package and a
series of scripts.  The master bias frame was created from all biases
and subtracted from all flats and science frames, and a linearity
correction was applied with {\tt IRLINCOR}, using cubic linearity
correction. Coefficients obtained from the INT Wide Field
Survey website\footnote{http://www.ast.cam.ac.uk/$\sim
$wfcsur/technical/ccd/} are listed in Table~\ref{linearity}.  Master
flats for each filter were created with appropriate bad pixel and
cosmic ray rejections, and the raw object frames were reduced to
science frames with {\tt CCDPROC}.

\begin{table}
\begin{minipage}{80mm}
\centering
\caption{Linearity Correction Coefficients for {\tt IRLINCOR}}
\label{linearity}
\begin{tabular}{c c r@{.}l r@{.}l}
\hline
CCD \# & $c_{1}$ & 
\multicolumn{2}{c}{ $c_{2}$ } & 
\multicolumn{2}{c}{ $c_{3}$ }\\
\hline
1 & 1.0 & -0&081918  &  0&012884 \\
2 & 1.0 & -0&0016384 & -0&0042947 \\
3 & 1.0 & -0&019660  &  0&0 \\
4 & 1.0 & -0&0049151 & -0&0021474 \\
\hline
\end{tabular}
\end{minipage}
\end{table}

\subsection{Image Difference analysis}
Lightcurves were derived from the science frame using difference image
analysis (DIA). The software used in this study has been described in 
\citet{bond01} and \citet{bramich05}, hereafter BBDIA. 
DIA attempts to match the point
spread function (PSF) between the frames in a time sequence by
generating a best seeing reference frame and then degrading that 
reference frame
to the seeing of the individual frames of the observing run. The
empirically-generated kernel solution models the changes to the PSF from one
image to another; it is solved for each image -- reference
pair. Non-varying stars leave no residual on the difference image, and
variables leave either positive or negative flux with respect to the
reference frame. The flux is compared with the reference frame and converted 
to magnitudes to generate light curves. DIA is {\it essential} in crowded
fields such as M31, where the high stellar density makes aperture
photometry inadequate, and varying backgrounds make PSF modelling
alone difficult. An overview of the procedure is given here, however, 
a more detailed description can be found in BBDIA.

\subsubsection{Photometry}
The reduced science frames were passed into the image subtraction
pipeline for reference frame generation, subtraction and photometry.

Reference frame generation was performed by selecting several frames 
with optimal
seeing and aligning and stacking them to increase signal to
noise. Around 12 frames were chosen to create the reference frame, all with a
FWHM of less than 1\,arcsec. Each science frame was aligned to the
reference frame and subtracted according to

\begin{equation}
\centering
Diff(x, y)=I(x, y)-Ref(x, y)\otimes Ker(u, v)-Bg(x, y)
\label{diaequation}
\end{equation}

where $Diff(x, y)$ is the difference image, $I(x, y)$ is any image in the time
series, $Ker(u, v)$ is the kernel that matches the image and reference
PSF, $Ref(x, y)$ is the reference frame and $Bg(x, y)$ is the spatially varying
differential background. The remaining frame consists of Poisson noise and the residuals 
of varying objects. Photometry can then be performed on each
residual flux to determine how it changes from frame-to-frame. An
IDL\footnote{IDL provided, under license, by Research Systems Inc.}
script then converted this flux to magnitudes relative to the stacked
reference frame according to equations~\ref{fluxtotal} and~\ref{magnitude}
(from BBDIA).

\begin{equation}
% f_{tot}(t)=10^{\frac{2}{5}\left(25.0-m_{ref}\right)}+\frac{f_{diff}(t)}{p(t)}
f_{tot}(t)=f_{ref} + \frac{f_{diff}(t)}{p(t)}
\label{fluxtotal}
\end{equation}
\begin{center}
\begin{equation}
m(t)=25.0-2.5~log(f_{tot}(t))
\label{magnitude}
\end{equation}
\end{center}

\begin{figure}

\caption{RMS diagrams for CCD\#1 of field A. Filters are $B$ (top) and
$V$ (bottom) respectively. The theoretical line is from {\tt SIGNAL},
available from the ING. At the faint end of the observations the 
elevation above the theoretical line is due to the sky brightness from 
unresolved background stars. At the bright end the deviation is caused 
by inadequacies in the model.  The unusual ``bump'' between magnitudes 
22 and 23 should be noted for the $V$ filter data.
This is may be caused by DIA
attempting to simulate the very faint and highly-crowded background
stars. As no EBs are identified at this magnitude, it is of no
relevance to this study but is worthy of note.
}
\includegraphics[height=7.5cm, angle=-90]{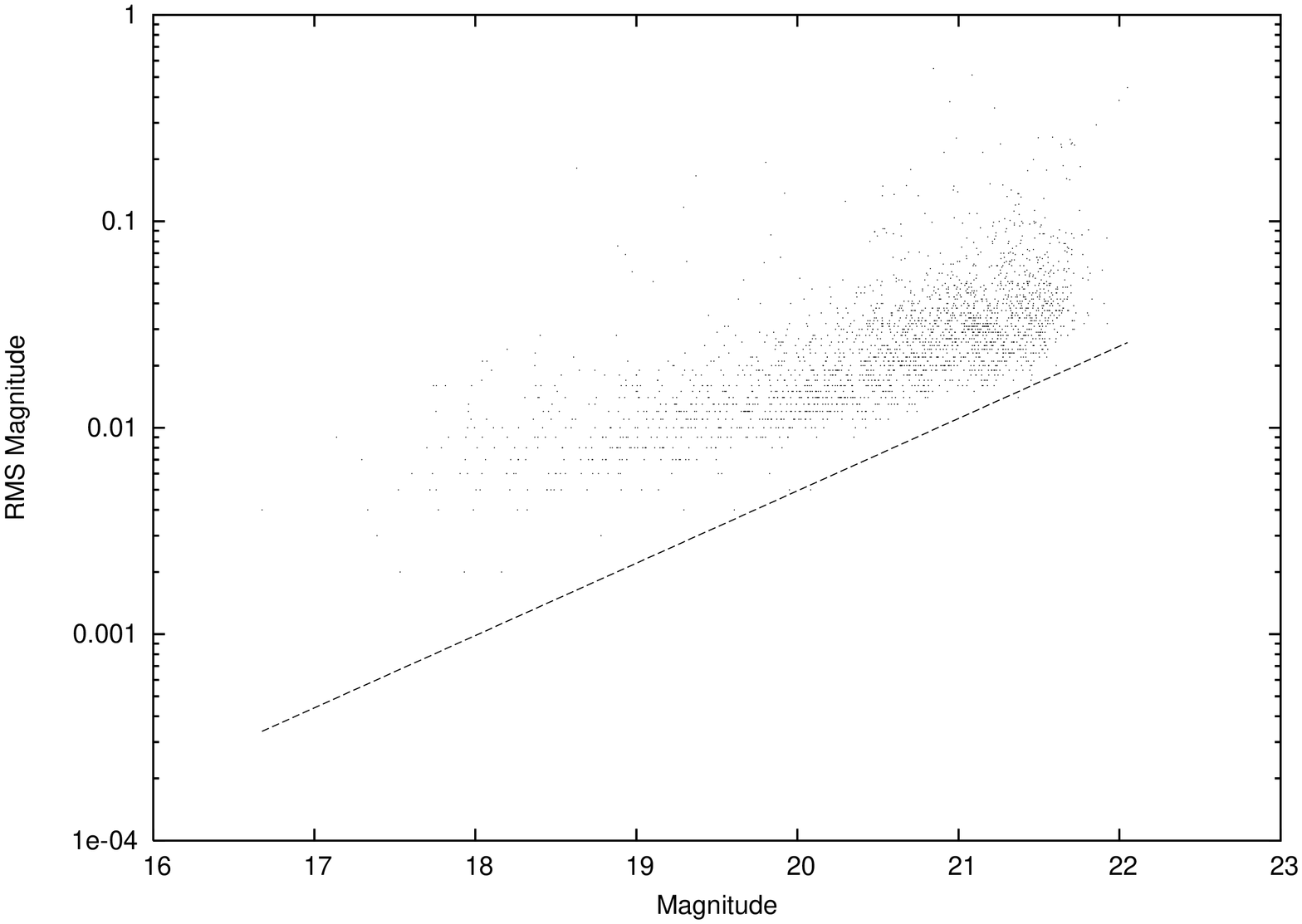}
\includegraphics[height=7.5cm, angle=-90]{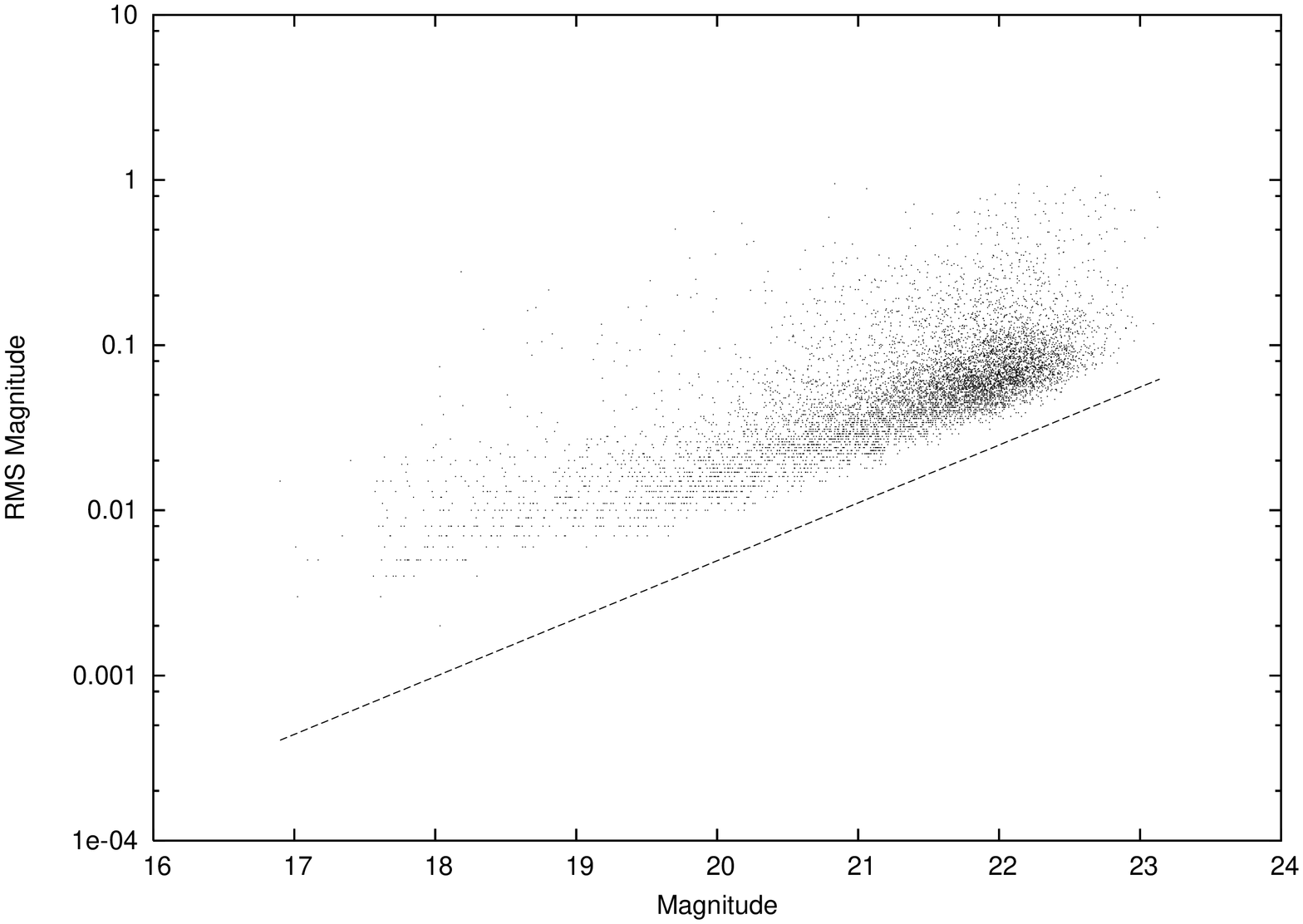}
\label{rmsdiag}
\end{figure}

The total flux, $f_{tot}$, from a star at time $t$ represents the
combination of reference flux $f_{ref}$ and difference flux $f_{diff}$. 
The photometric scale
factor $p(t)$ takes into account the change in extinction and exposure
time from frame to
frame. Uncertainties in all flux measurements are propagated
accordingly.  RMS diagrams for field A are shown in
Figure~\ref{rmsdiag}, and demonstrate a photometric accuracy of better
than 1 per cent for objects as faint as $21^{\rm st}$ magnitude.

The instrumental magnitudes were then mapped to the \citet{mag96} or
\citet{moch01} dataset via a linear transformation using the stars found
on the reference frame and solved for using a least squares process. 
Unresolved stellar blends can affect the absolute magnitude and 
amplitude calibrations.
The impact of such blends on the light curves is discussed in more
detail in Section~\ref{discussion}.

\subsubsection{Astrometry}
The astrometric solution was calculated for the frame using tasks from
the IRAF package {\tt IMCOORDS}. Once again, the dataset used for
reference frame was that of \citet{mag96}. Ten widely-spaced stars on the 
frame 
were used as a starting point for the astrometric solution, 
followed by the use of automated scripts to correlate positions of bright 
stars with entries in the USNO-B1.0\footnote{Provided by
United States Naval Observatory, Washington D.C.}  catalogue. This
procedure quantifies the pin-cushion distortion present on the
WFC; typical differences were approximately 0.4\,arcsec RMS, due to
a combination of the CCD distortion and that introduced in
constructing the reference frame.

The plate solution was then calculated for all the variables, converting 
frame positions to J2000.0 celestial coordinates in the equatorial
system based on the astrometric solution.

\subsubsection{Variability detection} 
Since DIA only selects the stars that are variable, the detection of 
variability is not an issue. However, the level of
variability is important so as to remove light curves that have low
intrinsic variability and high photometric error. An implementation of
the Stetson Variability Index (SVI), based on that adopted by
\citet{kaluzny98}, was employed to detect high amplitude stars with low
errors. The index is given by

\begin{center}
\begin{equation}
J = \frac{\sum_{k=1}^{n} w_{k} sgn(P_{k}) \sqrt{{P_{k}}}} {\sum^{n}_{k=1} w_{k} }
\label{stetsonindex}
\end{equation}
\end{center}

\noindent where k pairs of observations are defined, each with weight
$w_{k}$. The value $P_{k}$ is defined as the product of normalize
residuals of the paired observations $i$ and $j$:

\begin{displaymath}
P_{k} = \left \{
 \begin{array}{ll}
  \delta_{i(k)}\delta_{j(k)} ,& \mbox{if $i(k)\ne~j(k)$} \\
  \delta_{i(k)}^{2}-1 ,& \mbox{if $i(k)=j(k)$}\\
 \end{array}
 \right .
\label{pkstetson} 
\end{displaymath}

Finally $\delta$ is the magnitude residual of a given observation from
the mean, given over $n$ observations in a passband by

\begin{center}
\begin{equation}
\delta = \sqrt{\frac{n}{n-1}} \frac{\nu-\bar{\nu}}{\sum_{\nu}}
\label{stetsonindex1}
\end{equation}
\end{center}

where $\nu$ is the magnitude. $J$ should tend to zero for a
non-variable star and be positive for a variable. Observations $i$ and
$j$ will be in different pass bands at (approximately) the same
epoch. If only one observation is made at a particular epoch, $i(k)=j(k)$.

In determining the variability index, the maximum time separation in
the same filter for two measurements to be considered a pair was
around 90 minutes. If two points were a pair then the weight given
was 1.0, otherwise 0.25. 

\subsection{Matched Filters}
Classification of the  variable stars was performed automatically by
matching theoretical curves to the observed folded light curve by least
squares fit. Two simple classifying light curves were chosen; EB and
Cepheid. The classification code sampled multiple parameter space in
terms of period and light curve shape, varying the amplitude
and depth of eclipses, and searched for periods the range 0.5 to 15\,d.
Secondary eclipse amplitudes adopted lay in the range 0.3 to 0.9 magnitudes 
and assumed to be at phase 0.5 (i.e. circular orbits).  Primary eclipse 
amplitudes were scaled to
the amplitude of the light curve. This allowed broad classification to 
be made but in many cases light curves were mis-identified due to sparse 
sampling. Some binaries were
missed because the light curves had higher than normal scatter or several 
outlying observations distorted the $\chi^{2}$ fit.

To resolve this, a simple code was written that searched each
light curve in the time domain for linear trends, subject to various
gradients and fit coefficients. This is a similar approach to the
'box-fitting' technique used to search for the short eclipses of
extrasolar planets. This particular approach could not be used here 
because the light curves are not continuously sampled over the
eclipses. Instead, searches for the 'characteristic fragments' of an
EB were made -- sharp, almost linear, ascents and descents around eclipse.  
This revealed at least twice as many EBs as the simple matched filter
algorithm, but also revealed many false positives, such as Cepheids.
All light curves classified as EBs were inspected by eye to validate
their classification, with rejection of those that were not believed
to be binaries.

\subsection{Period Determination}
\begin{figure}
\begin{minipage}{80mm}
\centering
\caption{Simulation of orbital phase coverage for our dataset. The
phase space [0, 1] is divided into 100 equal bins. A dataset with at
least one point in each bin is said to have 100 per cent phase
coverage.}  
\includegraphics[height=7.5cm,angle=-90]{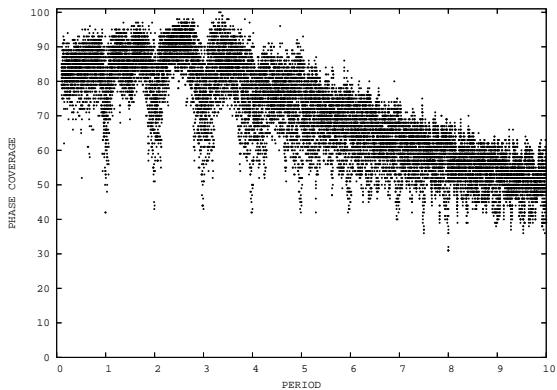}
\label{phasecoverage}
\end{minipage}
\end{figure}

Period finding in these sparse time series was difficult.  Various
methods such as PDM \citep{sw78}, ANOVA due to \citet{scz89} and
various string length techniques were tried; in general PDM gave more
consistent results. A visual inspection code was also drawn up,
whereupon a period was specified and the light curve folded appropriately.
In cases where automated period-finding techniques failed to give reliable 
results, variable incremental shifts in period were applied
and the folded light curve was inspected in real-time by eye.
The major difficulty in the period determination is aliasing. Due to the 
sampling of the data induced by the observational constraints, some of the
fainter and noisier light curves have several PDM peaks. In cases where the 
true peak could not be identified, it is impossible to say for certain 
whether the selected period is {\it definitely} the correct one without 
additional data, only that it is close; such cases have been noted in the 
tables. Ambiguity can also
arise when limited phase coverage prevents both primary and secondary
minima from being detected; a phase coverage simulation for this
dataset is shown in Figure~\ref{phasecoverage}.

\subsection{Results}

The EBs identified in our analysis, with periods (where available) are
listed in Tables~\ref{goodbinaries1}, \ref{goodbinaries2},
\ref{goodbinaries3} and~\ref{goodbinaries4}. In these tables, objects
with significant aliasing issues are annotated with their corresponding
periods.  Difference fluxes are an estimation of the depth of the primary 
and secondary eclipses in ADU/s, but in many cases the deepest part of 
the eclipses were not sampled by our data.
Quadrature magnitudes are listed, but
because of the likelihood of stellar blends affecting many objects
on the reference frame, the presented magnitudes are likely to be 
unreliable. 

Attempts have been made to match this catalogue with those of
others, in particular, the DIRECT Project and in some cases those of
\citet{bhgz88}, hereafter BHGZ88. Detached eclipsing binaries (DEB)
detected in this study and by the DIRECT Project are noted. 

\begin{table*}
\begin{minipage}{140mm}
\begin{center}
\caption{EBs in Field 1 of the Andromeda Galaxy. $D_{p}$ and $D_{s}$ are the depth of the primary and secondary eclipses
in ADU/s respectively.}
\begin{tabular}{lccllclll}
\hline
ID        & RA (J2000)  & DEC (J2000)  & P (d)           & $B_{max}$ & $V_{max}$ & $D_{p}$ & $D_{s}$ & Notes \\
\hline
f1BEB1448 & 0:44:29.272 & 41:23:01.446 & 3.1690          & 19.20     & 19.28     &  50     & 42      & V10550 DIRECT\\ 
f1BEB1500 & 0:44:26.974 & 41:23:42.531 & 3.3461          & 21.11     & 20.97     &  7.2    & 4.1     & 1, V9904 DIRECT\\ 
f1BEB1141 & 0:44:40.021 & 41:26:49.084 & 1.2614, 1.2636  & 21.55     & 21.74     &  2.1    &  0.9    & \\ 
f1BEB1205 & 0:44:37.988 & 41:29:23.819 & 3.5497          & 19.17     & 19.22     &  56     &  42     & V12650 DIRECT\\ 
f1BEB1598 & 0:44:48.675 & 41:29:15.557 & 5.7526          & 19.25     & -         &  27     & 27      & DEB, V9037 DIRECT\\
f1BEB2228 & 0:43:50.771 & 41:21:51.546 & 2.1767          & 20.36     & 20.34     &  12     & 8.5     & \\ 
f1BEB2065 & 0:43:59.846 & 41:21:20.108 & 2.7559          & 21.37     & 21.30     &  6.2    & 3.2     & \\ 
f1BEB2349 & 0:43:46.653 & 41:23:01.230 & 2.3705, 2.3783  & 20.47     & 20.55     &  13     & 12      & \\ 
f1BEB294  & 0:45:00.513 & 41:31:39.543 & 5.2495          & 19.60     & 19.65     &  14     & 13      & V6105 DIRECT, DEB? \\
f1BEB575  & 0:44:53.782 & 41:31:11.045 & 7.00            & 20.31     & 20.07     &  16     & 14      & V4903 DIRECT\\
f1BEB925  & 0:44:45.269 & 41:28:00.336 & 11.542          & 18.77     & 18.68     &  50     & 30      & V13944 DIRECT\\
f1BEB716  & 0:44:50.091 & 41:28:07.222 & 2.8724          & 20.17     & 20.32     &  22     & 11      & V14662 DIRECT\\ 
f1BEB1260 & 0:44:36.090 & 41:29:19.592 & 2.0429          & 19.81     & 19.92     &  16     & 11      & V12262 DIRECT\\ 
f1BEB1854 & 0:44:12.704 & 41:31:08.555 & 1.7461          & 21.00     & 21.02     &  7.5    & 4.9     & \\ 
f1BEB1813 & 0:44:13.710 & 41:22:07.134 & 2.7121          & 20.83     & 20.78     &  14     & 7.2     & \\ 
f1BEB417  & 0:44:57.387 & 41:30:07.021 & 2.1849          & 21.24     & 21.10     &  8.2    &  6.2    & \\ 
f1BEB2188 & 0:43:51.972 & 41:23:15.402 & 2.6264          & 21.27     & 21.24     &  4.5    &  3.5    & \\ 
f1BEB1152 & 0:44:39.373 & 41:25:59.551 & 2.1423          & 20.43     & 20.55     &  5.5    &  5.5    & DEB?\\ 
f1BEB1215 & 0:44:37.662 & 41:29:46.339 & 2.3009          & 20.40     & 20.43     &  12     &  10     & 1, V12594 DIRECT\\ 
f1BEB304  & 0:45:00.121 & 41:31:09.107 & 0.8577          & 21.20     & 21.28     &  4.0    &  3.5    & \\ 
f1BEB1429 & 0:44:29.889 & 41:23:27.589 & 2.3048          & 20.34     & 20.58     &  17     &  10.5   & V10732 DIRECT\\ 
f1BEB1407 & 0:44:30.758 & 41:24:09.575 & 1.8129          & 20.83     & 20.78     &  4.7    &  1.5     & \\ 
f1BEB2357 & 0:43:46.740 & 41:29:51.666 & 3.1646          & 20.61     & 20.67     &  11.4   &  10.4    & \\ 
f1BEB939  & 0:44:44.929 & 41:28:03.327 & 2.3893          & 21.17     & 21.03     &  6.0    &  5.0     & DEB\\ 
f1BEB788  & 0:44:48.572 & 41:27:27.870 & 7.1565, 7.2290  & 21.02     & 20.95     &  4      &  3      & 2, V14439 DIRECT\\ 
f1BEB467  & 0:44:56.477 & 41:30:37.289 & 2.0145          & 20.21     & 20.24     &  8      &  8      & \\ 
f1BEB1310 & 0:44:33.796 & 41:25:22.180 & 1.7353          & 21.16     & 21.39     &  4.3    & 4.3     & \\ 
f1BEB1262 & 0:44:35.768 & 41:24:41.847 & 2.6606          & 21.43     & 21.38     &  3      & 2       & \\ 
f1BEB989  & 0:44:43.604 & 41:26:34.587 & 1.6294          & 20.48     & 20.72     &  9      & 9       & \\ 
f1BEB207  & 0:45:05.051 & 41:31:56.579 & 1.6791          & 21.42     & 21.48     &   2.8   & 2.8     & \\ 
f1BEB1491 & 0:44:27.364 & 41:24:17.867 & 7.775           & 20.62     & 20.62     &   8.8   & 5.0     & \\ 
f1BEB452  & 0:44:56.829 & 41:31:11.271 & 4.258           & 20.36     & 20.26     &   8.0   & 8.0     & V5443 DIRECT\\ 
f1BEB321  & 0:44:59.345 & 41:30:45.767 & 2.6684          & 20.14     & 20.06     &  9.0    &  8.0    & BHGZ88?\\ 
f1BEB504  & 0:44:55.912 & 41:31:10.929 & 2.8987          & 20.30     & 20.32     &  6.1    &  6.0    & \\ 
f1BEB204  & 0:45:05.077 & 41:30:21.867 & 5.9234          & 21.37     & 21.18     &  -      & -       & 2, DEB\\ 
f1BEB732  & 0:44:49.943 & 41:28:50.896 & 3.8839          & 20.44     & 20.60     &  19     &  10     & V14653 DIRECT\\ 
f1BEB1748 & 0:44:16.667 & 41:24:19.373 & 3.3574          & 21.37     & 21.43     &  5.8    & 3.1     & \\ 
f1BEB1766 & 0:44:15.946 & 41:22:09.465 & 2.0834          & 21.36     & 21.40     &  5.4    & 3.5     & \\ 
f1BEB1181 & 0:44:38.179 & 41:25:30.335 & 2.4135          & 21.26     & 21.07     &  5.2    & 5.2     & DEB\\ 
\hline
\end{tabular}
\label{goodbinaries1}
\end{center}
\footnotemark[1]{Multiple aliases within $\pm$0.05\,d of the quoted period.} \\
\footnotemark[2]{Eclipse has no bottom, therefore flux depth unreliable}\\
\end{minipage}
\end{table*}

\begin{table*}
\centering
\begin{minipage}{140mm}
\begin{center}
\caption{EBs in Field 2 of the Andromeda Galaxy. }
\begin{tabular}{lccllcllll}
\hline
ID        & RA (J2000)  & DEC (J2000)  & P (d)           & $B_{max}$ & $V_{max}$ & $D_{p}$ & $D_{s}$ & Notes \\
\hline
f2BEB258  & 0:43:00.951 & 41:42:59.781 & 1.1793          & 21.24     & 21.30     &  5.7    & 5.5     & \\ 
f2BEB775  & 0:42:49.399 & 41:41:27.051 & 1.3603          & 21.22     & 21.21     &  3.8    & 2.7     & \\ 
f2BEB3137 & 0:42:47.357 & 41:32:55.285 & 0.80186         & 21.48     & 21.67     &  3.2    & 3.1     & \\ 
f2BEB450  & 0:43:29.063 & 41:42:31.489 & 1.9751          & 21.09     & 20.45     &  6.25   & 6.1     & Nearby HII region$^{1}$\\ 
f2BEB3693 & 0:43:13.561 & 41:30:00.140 & 1.7647          & 21.75     & -         &  4.5    & 1.5     & \\ 
f2BEB1197 & 0:42:55.767 & 41:40:12.860 & 0.83611         & 20.83     & 20.88     &  5.8    & 5.6     & \\ 
f2BEB330  & 0:43:25.453 & 41:42:50.649 & 1.1306          & 21.31     & 21.52     &  3.5    & 3.5     & \\ 
f2BEB3163 & 0:43:23.022 & 41:32:46.434 & 1.1595          & 22.10     & 21.17     &  2.0    & 1.7     & \\ 
f2BEB3695 & 0:43:31.679 & 41:29:59.588 & 1.6121          & 21.80     & 21.63     &  2.4    & 1.2     & \\ 
f2BEB563  & 0:43:06.584 & 41:42:11.405 & 3.0198          & 21.36     & 21.51     &  5.4    & 2.1     & \\ 
f2BEB1649 & 0:43:07.370 & 41:38:39.269 & 1.8272          & 22.18     & -         &  1.5    & 1.0     & DEB\\ 
f2BEB638  & 0:42:49.736 & 41:41:56.432 & 1.0639          & 22.94     & -         &  1.1    & 1.1     & \\ 
f2BEB5400 & 0:43:40.511 & 41:23:05.440 & 1.3987          & 21.47     & -         &  2.5    & 1.5     & \\ 
f2BEB162  & 0:43:27.209 & 41:43:12.934 & 2.7326          & 21.92     & 21.47     &  1.0    & 0.8     & \\ 
f2BEB3785 & 0:43:38.476 & 41:29:37.711 & 1.7114          & 21.33     & 21.47     &  5.8    & 3.5     & \\ 
f2BEB5221 & 0:42:49.977 & 41:23:37.359 & 1.8633          & 20.57     & 19.58     &  8.2    & 3.4     & \\ 
f2BEB1850 & 0:43:02.570 & 41:37:58.881 & 2.7955          & 22.33     & -         &  2.6    & 2.4     & DEB\\ 
f2BEB2027 & 0:42:48.376 & 41:37:25.345 & 1.5466          & 21.76     & 21.83     &  1.8    & 1.8     & \\ 
f2BEB2138 & 0:42:55.146 & 41:37:10.325 & 2.5179, 2.5176  & 21.76     & 21.40     &  2.5    & 1.7     & \\ 
f2BEB5325 & 0:43:40.248 & 41:23:21.349 & 3.2248, 3.2535  & 21.43     & 21.30     &  4.7    & 2.3     & \\ 
f2BEB2090 & 0:42:59.898 & 41:37:17.721 & 3.5830          & 20.62     & 20.53     &  7.5    & 3.7     & \\ 
f2BEB1756 & 0:42:59.288 & 41:38:16.991 & 5.5916          & 20.37     & 20.29     &  11     & 5.5     & \\ 
f2BEB3229 & 0:43:36.559 & 41:32:29.989 & 2.022           & 22.16     & 21.50     &  2.0    & 1.0     & \\ 
f2BEB630  & 0:42:46.505 & 41:41:57.627 & 5.0963          & 20.76     & 20.81     &  10     &   3.1   & \\ 
f2BEB2650 & 0:42:59.761 & 41:35:17.018 & 3.6682          & 21.04     & 20.89     &  3.6    &   3.6   & \\ 
f2BEB31   & 0:43:33.406 & 41:43:32.900 & 1.8884          & 22.00     & -         &  3.0    &   1.7   & \\ 
f2BEB622  & 0:43:31.104 & 41:42:03.049 & 2.9382          & 21.59     & 21.55     &  3.5    &   1.0   & \\ 
f2BEB5712 & 0:43:41.098 & 41:22:09.529 & 3.8868          & 21.15     & -         &  5.5    &  2.5    & \\ 
f2BEB5393 & 0:43:36.168 & 41:23:08.147 & 2.2315          & 21.50     & -         &  2.5    &  1.0    & \\ 
f2BEB986  & 0:43:29.868 & 41:40:52.399 & 5.989           & 22.26     & -         &  2.3    &  2.0    & \\ 
f2BEB393  & 0:43:31.551 & 41:42:40.693 & 1.6741          & 22.11     & 21.96     &   0.5   &  0.4   & \\ 
f2BEB1806 & 0:43:04.932 & 41:38:07.633 & 2.619, 2.609    & 21.94     & 21.72     &   1.5   &  0.5   & \\ 
f2BEB2857 & 0:43:00.719 & 41:34:16.919 & 2.0887          & 20.79     & 20.97     &   3.3   &  1.9    & \\ 
f2BEB371  & 0:43:35.647 & 41:42:44.426 & 2.8777          & 21.44     & 21.06     &   2.5   &  2.5    & \\ 
f2BEB1878 & 0:43:03.565 & 41:37:52.247 & 5.41            & 20.56     & 20.32     &   6.5   &  -      & 2 \\ 
f2BEB1352 & 0:42:53.333 & 41:39:42.250 & 4.61, 5.86      & 21.78     & 21.46     &   1.8   &  -      & 2 \\ 
f2BEB1227 & 0:42:46.247 & 41:40:06.210 & 3.264, 4.444    & 22.69     & -         &  1.5    &    -    & 2 \\ 
f2BEB629  & 0:43:26.095 & 41:42:01.193 & 4.2745          & 20.44     & 20.09     &  4.5    &  4.0    & \\ 
f2BEB4356 & 0:42:53.182 & 41:27:14.506 & 1.8776          & 19.92     & 20.17     &  5.5    &  4.0    & \\
\hline
\end{tabular}
\label{goodbinaries2}
\end{center}
\footnotemark[1]{from \citet{wb92a}}\\
\footnotemark[2]{Eclipse has no bottom, therefore flux depth unreliable}\\
\end{minipage}
\end{table*}

\begin{table*}
\centering
\begin{minipage}{140mm}
\begin{center}
\caption{EBs in Field 3 of the Andromeda Galaxy. }
\begin{tabular}{lccllcllll}
\hline
ID        & RA (J2000)  & DEC (J2000)  & P (d)                 & $B_{max}$ & $V_{max}$ & $D_{p}$ & $D_{s}$ & Notes \\
\hline
f3BEB2506 & 0:43:58.089 & 41:50:28.572 & 2.7491                &  20.53    & 20.45     &  8.1    &  5.6    & \\ 
f3BEB670  & 0:45:19.706 & 41:45:05.364 & 7.061                 &  20.38    & 19.41     &  -      &  -      & 2, V5407 DIRECT\\
f3BEB1265 & 0:44:50.255 & 41:51:24.232 & 1.9428                &  20.65    & 20.55     &  8.4    &  6.0    & \\ 
f3BEB1285 & 0:44:49.119 & 41:52:52.890 & 2.6270                &  19.52    & 19.52     &  20     &  20     & \\ 
f3BEB2388 & 0:44:04.561 & 41:48:52.159 & 2.6475                &  20.60    & 20.17     &  11     &  10     & \\ 
f3BEB1784 & 0:44:30.461 & 41:52:04.393 & 6.2241                &  20.17    & 20.73     &  14     &  13     & \\ 
f3BEB750  & 0:45:20.086 & 41:45:04.227 & 1.6041                &  20.13    & 20.45     &  13     &  10     & V4741 DIRECT\\
f3BEB456  & 0:45:32.433 & 41:47:42.806 & 2.7879                &  19.90    & 19.89     &  18     &  10     & V7393 DIRECT\\ 
f3BEB1778 & 0:44:30.629 & 41:51:56.121 & 1.4067                &  20.46    & 20.32     &  3.8    &  3.8    & \\ 
f3BEB760  & 0:45:19.706 & 41:45:05.364 & 5.75                  &  19.24    & 19.41     &  27     &  20     & DEB,V4636 DIRECT\\ 
f3BEB630  & 0:45:25.503 & 41:45:04.005 & 5.010                 &  20.04    & 19.99     &  8.5    &   -     & 2, V5912 DIRECT\\ 
f3BEB1547 & 0:44:37.084 & 41:52:25.119 & 4.768                 &  19.86    & 19.99     &    -    &   9     & 2\\ 
f3BEB251  & 0:45:38.924 & 41:47:49.825 & 1.2772                &  20.46    & 20.86     &   5.5   &   5.5   & \\ 
f3BEB2260 & 0:44:09.540 & 41:47:04.875 & 2.7721, 3.129, 3.012  &  20.77    & 20.96     &   4.5   &   -     & 2\\ 
f3BEB562  & 0:45:28.366 & 41:44:24.375 & 3.080                 &  20.54    & 20.56     &   6.0   &  6.0    & V6450 DIRECT\\ 
f3BEB580  & 0:45:28.150 & 41:49:32.961 & 4.37, 3.60            &  20.80    & 20.76     &   -     &    -    & 2, V6527 DIRECT\\ 
f3BEB651  & 0:45:24.220 & 41:46:26.850 & 6.798                 &  20.10    & 20.28     &   9     &   6.5   & \\ 
f3BEB505  & 0:45:31.116 & 41:46:49.598 & 4.504                 &  21.31    & 21.47     &   1.8   &   1.8   & \\ 
f3BEB1276 & 0:44:49.372 & 41:52:17.821 & 4.771                 &  20.47    & 20.39     &   -     &    -    & 2\\ 
\hline
\end{tabular}
\label{goodbinaries3}
\end{center}
\footnotemark[2]{Eclipse has no bottom, therefore flux depth unreliable}\\
\end{minipage}
\end{table*}

\begin{table*}
\centering
\begin{minipage}{140mm}
\begin{center}
\caption{EBs in Field 4 of the Andromeda Galaxy. }
\begin{tabular}{lccllcllll}
\hline
ID        & RA (J2000)  & DEC (J2000)  & P (d)                 & $B_{max}$ & $V_{max}$ & $D_{p}$ & $D_{s}$ & Notes \\
\hline
f4BEB1695 & 0:44:27.199 & 41:36:08.223 & 4.5186                & 20.19     & 20.03     &         &         & V1266 DIRECT\\ 
f4BEB821  & 0:45:12.519 & 41:37:26.321 & 2.3584                & 19.33     & 19.29     &         &         & V7940 DIRECT\\ 
f4BEB1157 & 0:45:05.377 & 41:33:40.442 & 1.7699                & 20.12     & 20.22     &         &         & V6846 DIRECT\\ 
f4BEB1180 & 0:45:04.843 & 41:37:29.350 & 3.0945                & 19.25     & 19.41     &         &         & V6840 DIRECT\\ 
f4BEB1962 & 0:44:11.074 & 41:34:08.197 & 2.9202                & 19.84     & 19.84     &         &         & \\ 
f4BEB2294 & 0:43:53.920 & 41:36:42.418 & 6.242                 & 20.89     & 21.02     &         &         & \\ 
f4BEB246  & 0:45:26.694 & 41:41:03.179 & 2.0833                & 20.71     & 20.95     &         &         & V6024 DIRECT\\ 
f4BEB1672 & 0:44:28.010 & 41:36:57.330 & 2.8949                & 21.62     & -         &         &         & \\ 
f4BEB1763 & 0:44:24.763 & 41:39:02.422 & 4.763                 & 19.98     & 20.04     &         &         & V888 DIRECT \\
f4BEB1802 & 0:44:22.332 & 41:38:51.216 & 0.2326                & 18.68     & 17.80     &         &         & V438 DIRECT, Local WUMa\\
f4BEB74   & 0:45:37.723 & 41:43:07.777 & 1.9302                & 20.94     & 20.81     &         &         & V8420 DIRECT\\ 
f4BEB930  & 0:45:10.294 & 41:36:47.119 & 6.0972, 3.049, 5.1395 & 18.78     & 18.85     &         &         & V7628 DIRECT\\ 
f4BEB1621 & 0:44:31.171 & 41:36:15.288 & 1.553                 & 20.52     & 19.61     &         &         & 1\\ 
f4BEB1893 & 0:44:15.941 & 41:37:29.369 & 2.7262                & 21.84     & -         &         &         & \\ 
f4BEB1168 & 0:45:05.200 & 41:38:46.464 & 0.91664               & 20.77     & 20.71     &         &         & V1555 DIRECT\\ 
f4BEB2290 & 0:43:54.185 & 41:37:13.150 & 3.335                 & 20.65     & 20.73     &         &         & \\ 
f4BEB920  & 0:45:10.432 & 41:36:34.421 & 6.7657                & 20.40     & 20.33     &         &         & Unusual LC\\ 
f4BEB990  & 0:45:09.190 & 41:38:41.867 & 4.4523                & 20.25     & 20.36     &         &         & \\ 
f4BEB296  & 0:45:24.691 & 41:39:41.341 & 2.4223                & 21.61     &           &         &         & \\ 
f4BEB1617 & 0:44:31.538 & 41:37:22.253 & 4.6471                & 21.76     &           &         &         & \\ 
f4BEB1994 & 0:44:10.634 & 41:39:49.699 & 5.4995                & 21.72     &           &         &         & 1\\ 
f4BEB527  & 0:45:18.676 & 41:40:47.040 & 2.0215                & 20.61     & 20.67     &         &         & \\ 
f4BEB1465 & 0:44:49.047 & 41:32:59.691 & 0.95183               & 21.28     & 21.26     &         &         & \\ 
f4BEB2006 & 0:44:10.277 & 41:39:49.409 & 2.8143                & 20.98     & 20.79     &         &         & \\ 
f4BEB2050 & 0:44:08.816 & 41:39:49.889 & 1.7957                & 21.33     &           &         &         & \\ 
f4BEB953  & 0:45:09.761 & 41:34:17.287 & 1.7806                & 21.59     & 21.28     &         &         & \\ 
f4BEB698  & 0:45:14.908 & 41:40:30.521 & 2.4350                & 20.49     & 20.47     &         &         & \\ 
f4BEB428  & 0:45:20.771 & 41:38:27.814 & 1.9108                & 21.40     & -         &         &         & \\ 
f4BEB453  & 0:45:20.351 & 41:41:44.412 & 1.0791                & 20.93     & 21.04     &         &         & \\ 
f4BEB1915 & 0:44:13.536 & 41:39:19.126 & 1.9272                & 21.02     & 20.91     &         &         & \\ 
\hline
\end{tabular}
\label{goodbinaries4}
\end{center}
\footnotemark[1]{Multiple aliases within $\pm$0.05\,d of the quoted period.} \\
\footnotemark[2]{Eclipse has no bottom, therefore flux depth unreliable}\\
\end{minipage}
\end{table*}

\section{Discussion}\label{discussion}
% ideas, writing will need to be more fluid

The matched filter analysis detected many objects with fragments of
light curves that are consistent with that of an eclipsing binary. In cases 
where sufficient data were available to determine periods, attempts were made  
to rule out aliases by searching for periods in subsets of the data to 
determine consistency in the power spectra, and by randomizing the brightness
measurements on the time axis to derive significance contours.

The DIRECT Project found 34 EBs coincident with the observed field, and the
observations tabulated in this study have recovered all but 5 of these. The
undetected objects are all long period systems, objects for which this study 
is not expected to be sensitive (see Fig.~\ref{phasecoverage}).  Of particular
interest are two objects, V6105B and V888B, classified as detached
systems by DIRECT.  Both of these objects have been recovered and their
detached nature confirmed. In principle, these objects could be used for 
accurate distance determinations, and DIRECT considers them suitable systems
for detailed follow-up both photometrically and spectroscopically
\citep{macri04}.

Other noteworthy systems include: \\
{\it f4BEB1802}. This is a short period system (P\,=\,0.232\,d),
originally discovered by DIRECT (\# V438). Its brightness and colour
suggest it is a foreground object. The photometry presented here
does not constrain the period well due to its high scatter when phased,
indicating long-term intrinsic variability.\\
{\it f1BEB1205, f1BEB1448} (DIRECT \#V10550), {\it f3BEB456}, and 
{\it f4BEB1180} are all bright semi-detached systems.\\
{\it f3BEB760}. This is a  bright, long-period detached system, and is
ideal for follow-up spectroscopy.  {\it f4BEB1763, f1BEB1181, f1BEB939}
and {\it f2BEB1850} are also detached systems. The photometry
presented here is not of sufficient quality for accurate light curve
analyses.

\subsection{Blending}

Unresolved background stars are believed to be a major cause of
elevated background levels in the images studied here; the effect of 
this problem is demonstrated in Figure~\ref{rmsdiag}. 
It is reasonable to expect that crowding issues will affect the accuracy of 
the photometry presented here due to a significant proportion of unresolved 
blended images. This problem of blending is discussed in detail in 
\citet{kiss05}. 
{\it Blending will contribute errors in the absolute flux
measurement on the reference image, and hence the quadratures
magnitudes and eclipse depths of blended EBs will be incorrect}. It
is for this reason that results are presented here in terms of difference flux
units, as periods and difference fluxes will be relatively unaffected
by blending. The level of blending can be estimated with high resolution
imaging by HST and such imaging of systems that are to be used in
distance determination is essential.

The median FWHM of the images analysed here, $\sim 1$\,arcsec, corresponds 
to a spatial resolution of 3.6\,pc in M31. \citet{moch04} have investigated
the third light contamination of ground-based photometry of 22
Cepheids in M31 using HST imaging. They find in the $V$-band that, on
average, $\sim$19 per cent of the stellar flux is caused by a
resolvable companion. Given that the physical size of a WFPC2 pixel is
$\sim$0.36\,pc in M31, and that a young cluster could be $<$0.1\,pc in
diameter, then clearly the \citet{moch04} estimate is a lower limit
(even without considering issues of binarity). The brightest M31
binaries discussed here have $M_V \sim$ --5.1 and therefore must have O-type
components. Assuming that a Salpeter IMF is applicable in M31
\citep{velt04}, one might expect that on average $\sim$7 B-stars could 
fall within a single HST pixel and contribute third light.

% dlp salpeter imf with gamma=-1.59, M(O-star)=60 Msol, M(B-star)=18 Msol
% N(O)/N(B)=0.147
% 
%At the distance of M31 (0.75 Mpc), 1 arcsec resolution (typical seeing conditions) corresponds
%to 3.6 pc. As it is likely all $O$ stars are actually born in clusters and that
%their short lifetimes suggest these associations should still exist. Therefore
%we have to face the real possibility of third light contamination in our
%light curves. With this data we can only place weak limits on the level
%of contamination.

The EBs with measured periods form a subset of those objects found. Where
light curves were sufficient to indicate a {\it probable} EB, there were
many cases where an estimate of the period was not possible. There
are around 160 of these objects spread across the four fields.  The level of
completeness in detection of the EBs is presumed to be low -- many
will not have been detected due to inadequate sampling.  Overall, 127
eclipsing binaries have been detected -- 98 of which are newly
discovered.  The matched filter analysis also detected many Cepheids
and other long-period variables. A discussion of these results will follow in
a subsequent paper.

\begin{figure*}
\centering
\begin{minipage}{170mm}
\caption{A selection of $B$-band light curves of interest from the dataset. The flux units
are adu/sec in each case.}
%\begin{center}
\includegraphics[height=150mm, angle=-90]{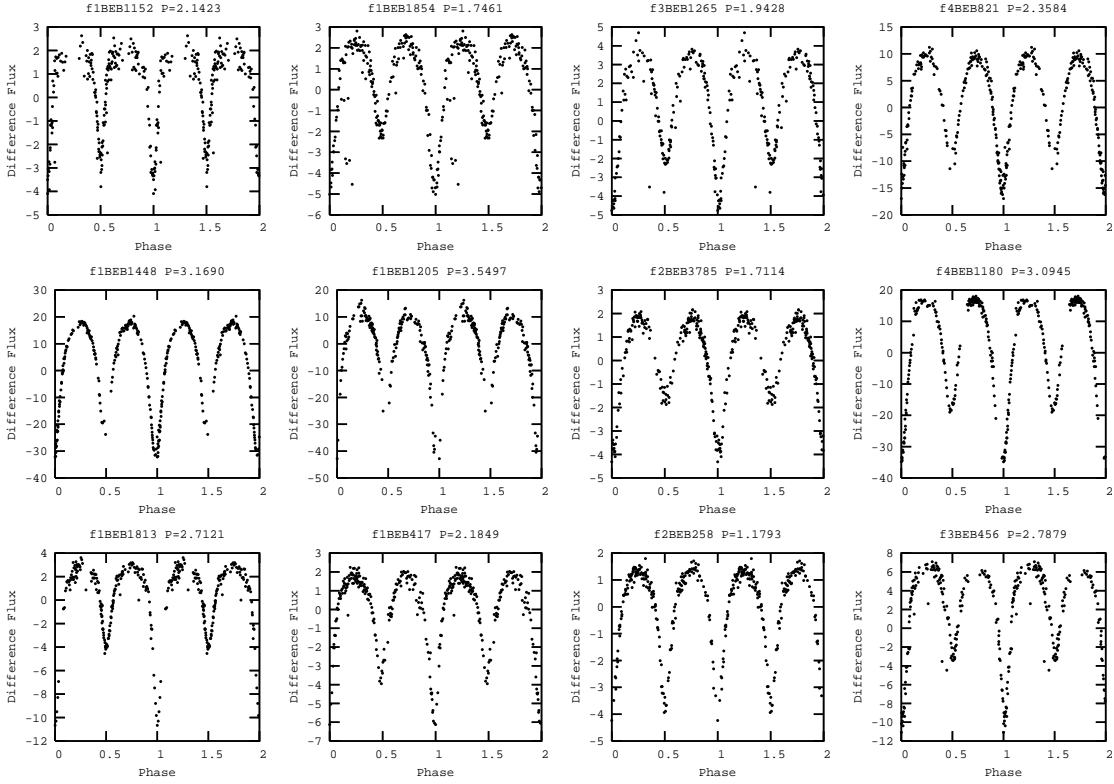}
%\end{center}
\end{minipage}
\label{goodlightcurves}
\end{figure*}

\begin{figure*}
\centering
\begin{minipage}{170mm}
\caption{$B$-band light curves of detached EBs. The flux units
are adu/sec in each case.}
%\begin{center}
\includegraphics[height=150mm,angle=-90]{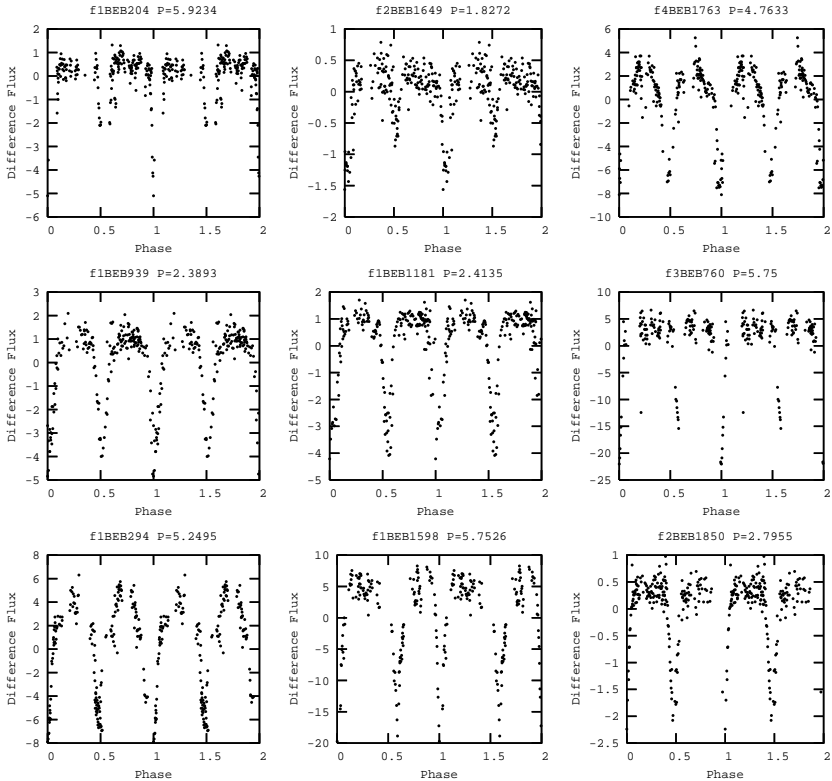}
%\end{center}
\end{minipage}
\label{detachedlightcurves}
\end{figure*}

\section*{Acknowledgements}
The INT is operated on the island of La Palma by the Isaac Newton
Group in the Spanish Observatorio del Roque de los Muchachos of the
Instituto de Astrofisica de Canarias The production of this work made
use of the CONDOR distributed computing software, the Aladin and Vizier
services and the NASA/ADS abstract service.

\bsp

\label{lastpage}


\begin{thebibliography}{}
 \bibitem[\protect\citeauthoryear{Andersen}{1991}]{andersen91}
 Andersen J., 1991, A\&A Rev. 3, 91.
% \bibitem[\protect\citeauthoryear{Alves et al.}{2004}]{alves04} Alves
% D.R., New Ast. Rev., 48, Issue 9, 659.
 \bibitem[\protect\citeauthoryear{Baade \& Swope}{1963}]{bs1963}
 Baade W., Swope H.H., AJ, 1963, 68, 435.
% \bibitem[\protect\citeauthoryear{Baade \& Swope}{1965}]{bs1965}
% Baade W., Swope H.H., AJ, 1965, 70, 212.
 \bibitem[\protect\citeauthoryear{Berkhuijsen et al.}{1988}]{bhgz88}
 Berkhuijsen E.M., Humphreys R.M., Ghigo F.D., Zumach W., 1988, A\&AS 76, 65.
% \bibitem[\protect\citeauthoryear{Bonanos et al.}{2003}]{bonanos03}
% Bonanos A.Z., Stanek K.Z., Sasselov D.D., Mochejska B.J.,
% Macri L.M., Kaluzny J., 2003. 
% \bibitem[\protect\citeauthoryear{Bonanos et al.}{2004}]{bonanos04}
% Bonanos A.Z, American Astronomical Society Meeting 205, \#73.04, 2004.
 \bibitem[\protect\citeauthoryear{Bond et al.}{2001}]{bond01}
 Bond I.A., Abe F., Dodd R.J., Hearnshaw J.,B., Honda M., Jugaku
 J., Kilmartin P.M., Marles A. et al., 2001, MNRAS 327, 868.
 \bibitem[\protect\citeauthoryear{Bramich et al. }{2005}]
{bramich05} Bramich D.M., Horne K., Bond I.A., Street R.A., 
Collier Cameron A., Hood B., Cooke J., James D., et al., 2005, MNRAS, 359, 1096.
 \bibitem[\protect\citeauthoryear{Clausen}{2003}]{clausen03} Clausen 
 J.V., 2003, A\&A, 402, 509.
 \bibitem[\protect\citeauthoryear{Clausen}{2004}]{clausen04} Clausen 
 J.V, 2004, New Astron. Rev. 48, Issue 9, 679.
 \bibitem[\protect\citeauthoryear{Clementini et al.}{2001}]{clem}
 Clementini G., Federici L., Corsi C., Cacciari C., Bellazzini M.,
 Smith H.A., 2001, ApJ, 559, L109.
 \bibitem[\protect\citeauthoryear{Cole}{1998}]{cole98}
 Cole A.A., 1998, ApJ, 500, L137.
 %\bibitem[\protect\citeauthoryear{Fitzpatrick et al.}{2002}]{fitzpatrick02}
 %Fitzpatrick E.L., Ribas I., Guinan E.F., DeWarf L.E., Maloney F.P., 
 %Massa D., 2002, ApJ, 564, 260.
 \bibitem[\protect\citeauthoryear{Fitzpatrick et al. }{2003}]
{fitzpatrick03} Fitzpatrick E.L., Ribas I., Guinan E.F., Maloney F.P., 
 Claret A., 2003, ApJ 587, 685.
 \bibitem[\protect\citeauthoryear{Freedman et al.}{2001}]{fm01}
 Freedman W.L., Madore B.F., Gibson B.K., Ferrarese L., Kelson D.D., 
 Sakai S., Mould J.R., Kennicutt R.C., et al., 2001, ApJ, 553, 47.
 \bibitem[\protect\citeauthoryear{Gaposhkin}{1968}]{gap68} Gaposhkin 
 S., 1968, PASP, 80, 556.
 \bibitem[\protect\citeauthoryear{Guinan}{1998}]{guinan98}
 Guinan E.F., Fitzpatrick E.L., Dewarf L.E., Maloney F.P., Maurone P.A.,
  Ribas I., Pritchard J.D., Bradstreet D.H., 1998, ApJ, 509, L21.
 \bibitem[\protect\citeauthoryear{Guinan}{2004}]{guinan04} Guinan 
 E., 2004, New Astron. Rev. 48, Issue 9, 647.
 \bibitem[\protect\citeauthoryear{Harries et al.}{2003}]
{harries03} Harries T.J., Hilditch R.W., Howarth I.D.,  2003, MNRAS, 339, 157.
% \bibitem[\protect\citeauthoryear{Hilditch}{2004}]{hilditch2004} Hilditch 
%  R.W., 2004, New Astron. Rev., 48, Issue 9, 687.
 \bibitem[\protect\citeauthoryear{Hilditch et al.}{2005}]
{hilditch2005} Hilditch R.W., Howarth I.D., Harries T.J.,  2005, MNRAS, 357, 304.
% \bibitem[\protect\citeauthoryear{Howarth et al.}{2004}]
%{hilditch04a} Hilditch R.W., Harries T.J., Howarth I.D., 2004, New Ast. Rev., 48,
% Issue 9, 687.
 \bibitem[\protect\citeauthoryear{Kaluzny et al.}{1998}]{kaluzny98} 
 Kaluzny J., Stanek K.Z., Krockenberger M., Sasselov D.D., Tonry J.L.,
  Mateo M., 1998, AJ, 115, 1016.
 \bibitem[\protect\citeauthoryear{Kiss \& Bedding}{2005}]{kiss05}
 Kiss L.L., Bedding T.R., 2005, MNRAS, 358, 883. 
 \bibitem[\protect\citeauthoryear{Macri}{2004}]{macri04} Macri L., 
 2004, New Ast. Rev., 48, Issue 9, 675.
 \bibitem[\protect\citeauthoryear{Magnier}{1996}]{mag96} Magnier E.A., 
 1996, A\&A Supp., 96, 379.
 \bibitem[\protect\citeauthoryear{McConnachie et al.}{2005}]
{mcc2005} McConnachie A.W., Irwin M.J., Ferguson A.M.N., 
 Ibata R.A., Lewis G.F., Tanvir N., 2005, MNRAS 356, 979.
 \bibitem[\protect\citeauthoryear{Mochejska et al.}{2001}]
{moch01} Mochejska B.J., Kaluzny J., Stanek K.Z., 
 Sasselov D.D., 2001, AJ 122, 1383.
 \bibitem[\protect\citeauthoryear{Mochejska et al.}{2004}]
{moch04} Mochejska B.J., Macri L.M., Sasselov D.D.,  Stanek K.Z., 
 Sasselov D.D., 2004, ASPC 310, 41.
 \bibitem[\protect\citeauthoryear{Mould et al.}{2004}]{mould04} 
 Mould J., Saha A., Hughes S., 2004, ApJS, 154, 623.
% \bibitem[\protect\citeauthoryear{Pritchard, J.D. et al.}{1998}]{pritchard98}
% Pritchard J.D., Tobin W., Clark M., Guinan E.F., 1998, MNRAS, 299,
% 1087.
 \bibitem[\protect\citeauthoryear{Ribas}{2002}]{ribas02} Ribas I.,
  Fitzpatrick E.L., Maloney F.P., Guinan E.F., Udalski A., 2002, ApJ, 574, 771.
% \bibitem[\protect\citeauthoryear{Ribas}{2004}]{ribas04} Ribas I.,
% Jordi C., Vilardell F., Giménez Á., Guinan E.F., 2004, New Ast.
% Rev., 48, Issue 9, 755.
 \bibitem[\protect\citeauthoryear{Schwarzenberg-Czerny}{1989}]{scz89}
 Schwarzenberg-Czerny A., 1989, MNRAS, 241, 153.
% \bibitem[\protect\citeauthoryear{Southworth}{2004}]{southworth04}
% Southworth J., Maxted P.F.L., Smalley B., 2004, MNRAS, 349, 547.
 \bibitem[\protect\citeauthoryear{Stellingworth}{1978}]{sw78}
 Stellingworth R.F., 1978, ApJ, 224, 953.
 \bibitem[\protect\citeauthoryear{Veltchev et al.}{2004}]
{velt04} Veltchev T., Nedialkov P., Borisov G., 2004, A\&A, 426, 495.
 \bibitem[\protect\citeauthoryear{Walterbos \& Braun}{1992}]{wb92a}
 Walterbos R.A.M, Braun R., 1992, A\&AS, 92, 625.
 \bibitem[\protect\citeauthoryear{Wilson}{2004}]{wilson04} Wilson 
 R.E., 2004, New Astron. Rev., 48, Issue 9, 695.
\end{thebibliography}
\end{document}